\documentclass[sigconf]{acmart}
\AtBeginDocument{%
  }

\copyrightyear{2026}
\acmYear{2026}
\setcopyright{cc}
\setcctype{by}
\acmConference[KDD '26]{Proceedings of the 32nd ACM SIGKDD Conference on Knowledge Discovery and Data Mining V.2}{August 09--13, 2026}{Jeju Island, Republic of Korea}
\acmBooktitle{Proceedings of the 32nd ACM SIGKDD Conference on Knowledge Discovery and Data Mining V.2 (KDD '26), August 09--13, 2026, Jeju Island, Republic of Korea}
\acmDOI{10.1145/3770855.3818475}
\acmISBN{979-8-4007-2259-2/2026/08}
\usepackage{enumitem}
\usepackage{array}
\usepackage{makecell}
\usepackage{multirow}
\usepackage{amsmath}
\usepackage{boldline} % for \Xhline

\begin{document}

\title{SaFRO: Satisfaction-Aware Fusion via Dual-Relative Policy Optimization for Short-Video Search}

\author{Renzhe Zhou}
\affiliation{%
  \institution{Kuaishou Technology}
  \city{Hangzhou}
  \country{China}}
\email{zhourenzhe03@kuaishou.com}
\authornote{Equal contribution.}

\author{Songyang Li}
\affiliation{%
  \institution{Kuaishou Technology}
  \city{Beijing}
  \country{China}}
\email{lisongyang03@kuaishou.com}
\authornotemark[1]

\author{Feiran Zhu}
\affiliation{%
  \institution{Kuaishou Technology}
  \city{Hangzhou}
  \country{China}}
\email{zhufeiran03@kuaishou.com}
%\authornotemark[2]
\authornote{Corresponding author.}

\author{Chenglei Dai}
\affiliation{%
  \institution{Kuaishou Technology}
  \city{Hangzhou}
  \country{China}}
\email{daichenglei@kuaishou.com}

\author{Yi Zhang}
\affiliation{%
  \institution{Kuaishou Technology}
  \city{Beijing}
  \country{China}}
\email{zhangyi49@kuaishou.com}

\author{Yi Wang}
\affiliation{%
  \institution{Kuaishou Technology}
  \city{Beijing}
  \country{China}}
\email{wangyi05@kuaishou.com}

\author{Jingwei Zhuo}
\affiliation{%
  \institution{Unaffiliated}
  \city{Beijing}
  \country{China}}
\email{zhuojw10@gmail.com}


\renewcommand{\shortauthors}{Renzhe Zhou et al.}

\begin{abstract}
Multi-Task Fusion plays a pivotal role in industrial short-video search systems by aggregating heterogeneous prediction signals into a unified ranking score. However, existing approaches predominantly optimize for immediate engagement metrics, which often fail to align with long-term user satisfaction. While Reinforcement Learning (RL) offers a promising avenue for user satisfaction optimization, its direct application to search scenarios is non-trivial due to the inherent data sparsity and intent constraints compared to recommendation feeds. To this end, we propose SaFRO, a novel framework designed to optimize user satisfaction in short-video search. We first construct a satisfaction-aware reward model that utilizes query-level behavioral proxies to capture holistic user satisfaction beyond item-level interactions. Then we introduce Dual-Relative Policy Optimization (DRPO), an efficient policy learning method that updates the fusion policy through relative preference comparisons within groups and across batches. Furthermore, we design a Task-Relation-Aware Fusion module to explicitly model the interdependencies among different objectives, enabling context-sensitive weight adaptation. Extensive offline evaluations and large-scale online A/B tests on Kuaishou short-video search platform demonstrate that SaFRO significantly outperforms state-of-the-art baselines, delivering substantial gains in both short-term ranking quality and long-term user retention.
\end{abstract}




\maketitle

\section{Introduction}
\begin{figure}[tbp]
  \centering
  \small
  \includegraphics[width=1.0\linewidth]{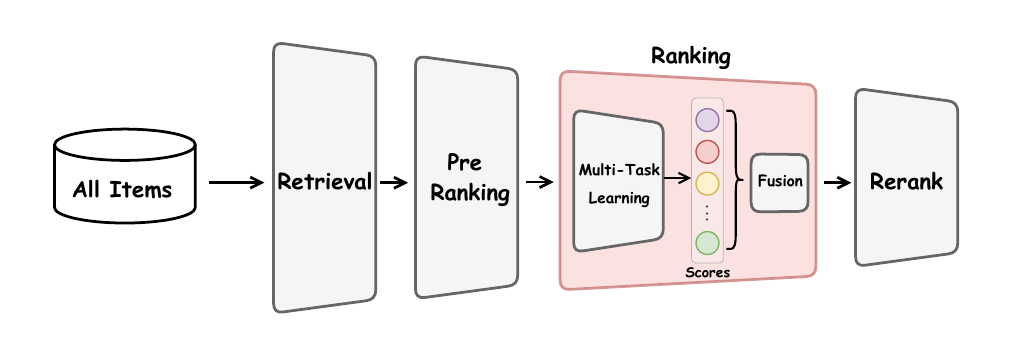} 
  %\vspace{-15pt}
  \caption{Cascaded search system architecture.}
  %\vspace{-5pt}
  \label{fig:rank}
\end{figure}

Industrial short-video search systems typically employ a cascaded architecture (as shown in Figure~\ref{fig:rank}): a retrieval stage first fetches candidate items from a large-scale index, followed by one or more ranking stages that progressively refine the order using richer features and more sophisticated models. The ranking pipeline commonly adopts a two-stage paradigm comprising Multi-Task Learning (MTL) and Multi-Task Fusion (MTF)~\cite{mtl_momma2022,cao2020ranking,chen2018gradnorm,ma2018modeling,su2024stem,yu2020gradient}. Specifically, the MTL model jointly predicts multiple objectives (e.g., click-through rate, long-play rate), while MTF synthesizes these heterogeneous signals into a final ranking score. Since the correlation between immediate engagement and long-term user satisfaction is fundamentally complex and obscure, MTF is essential for scalarizing multi-objective predictions into a unified metric that aligns with user satisfaction.

Reinforcement learning (RL)~\cite{sutton1998reinforcement} has emerged as a promising framework for multi-objective optimization in recommendation systems, primarily due to its ability to handle delayed reward signals and support long-horizon objectives~\cite{batchrl_mtf, rlur, pei2019value,chen2024cache,zhang2024unex}. Compared with supervised learning methods that mainly optimize immediate feedback, RL naturally models sequential user--system interactions and can balance short-term engagement with long-term user value. In short-video platforms like TikTok and Kuaishou, interactions in recommendation scenario (e.g., feed) generate dense and sequential user feedback, which makes RL particularly well-suited for optimizing cumulative outcomes such as revisit behavior or user retention~\cite{rlur,auro}. However, directly applying RL solutions from recommendation to search is non-trivial due to two fundamental differences:

%Reinforcement learning (RL) ~\cite{sutton1998reinforcement} has emerged as a promising framework for multi-objective optimization in recommendation systems, primarily due to its ability to handle delayed reward signals and support long-horizon objectives~\cite{batchrl_mtf, rlur, pei2019value,chen2024cache,zhang2024unex,pei2019value}. In short-video platforms like TikTok and Kuaishou, interactions in recommendation scenario (e.g., feed) generate dense and sequential user feedback, which makes RL particularly well-suited for optimizing cumulative outcomes such as revisit behavior or user retention~\cite{rlur,auro}. However, directly applying RL solutions from recommendation to search is non-trivial due to two fundamental differences:

\textbf{(i) Intent and relevance constraints.} Search is driven by explicit queries that express well-defined user intent and impose a strong relevance constraint. Recommendation, by contrast, typically lacks an explicit intent signal and serves latent, exploratory preferences, where diversity and exploration are often encouraged rather than constrained.

\textbf{(ii) Output form and interaction dynamics.} Search returns a page-level ranked list, where users aim to find a specific answer or target item and often terminate quickly once satisfied, making a session closely aligned with a single request. Recommendation delivers an infinite scroll content stream, promoting continuous consumption and requests within a session.

%\begin{table}[t]
%\centering
%\caption{Positive ratio of immediate feedback in search and recommandation sceniro. Click in recommandation represents valid play.}
%\label{tab:se_rec}
%\small
%\renewcommand{\arraystretch}{1.2}
%\begin{tabular}{c|c|cccccc}
%\Xhline{1pt}
%Sceniro & Items & Click & Long-play & Like & Follow   \\
%\hline
%Search & 3010195 & 25.87\% & 10.64\% & 0.62\% & 0.08\% \\
%Recommandation & 3054895 & 54.00\% & 40.87\% & 2.62\% & 0.16\% \\
%\hline
%\end{tabular}
%\end{table}

Consequently, search generates significantly sparser interaction data compared to recommendation: users typically inspect only a few results per page and provide limited feedback before ending their session. For instance, on the Kuaishou short-video platform, the density of positive interactions (e.g., clicks, long-plays, and likes) in search is approximately one-tenth of those observed in recommendation feeds. Furthermore, the length of user search behavior sequences is roughly two orders of magnitude shorter than that of recommendation sequences. This data sparsity makes it challenging to attribute long-term outcomes to specific ranking decisions, posing a fundamental obstacle to directly optimizing user satisfaction in search. Therefore, we argue that effective MTF in short-video search must go beyond immediate interactions and explicitly incorporate delayed signals of user satisfaction.

To this end, we propose SaFRO (\textbf{Sa}tisfaction-aware \textbf{F}usion via dual-\textbf{R}elative policy \textbf{O}ptimization), a multi-task fusion framework tailored for short-video search. SaFRO leverages query-level signals together with user retention to construct a unified satisfaction-aware reward model. To enable efficient policy learning under sparse and delayed feedback, we propose Dual-Relative Policy Optimization (DRPO) that updates the policy through relative comparisons both within groups and across batches. Furthermore, to explicitly model interdependencies among different objectives, SaFRO incorporates a Task-Relation-Aware Fusion module to generate context-sensitive logit adjustments for adaptive trade-offs.

Our contributions are summarized as follows:  
\begin{itemize}
    \item To the best of our knowledge, we are the first to model user satisfaction for industrial short-video search, unifying query-level signals and retention into a satisfaction-aware model for optimizing long-term user satisfaction.
    \item We propose Dual-Relative Policy Optimization, which employs dual-relative advantage for policy training, together with a composite reward that jointly incorporates engagement, satisfaction, and format constraints.  
    \item We introduce a Task-Relation-Aware Fusion module that explicitly models inter-task relationships via a learnable task-relation matrix, producing context-adaptive logit adjustments for dynamic multi-objective balancing.
\end{itemize}

\section{Related Work}
\textbf{Multi-Task Learning}. MTL has become a prevalent paradigm in industrial ranking systems, enabling joint prediction of diverse user feedback signals such as clicks, likes, watch time, and conversions~\cite{mtl_momma2022,cao2020ranking}.
By sharing representations across related tasks, MTL improves data efficiency and generalization compared to single-task models.
Early approaches like Shared-Bottom architecture~\cite{caruana1997multitask} and Cross-Stitch networks~\cite{misra2016cross} explored various feature sharing mechanisms.
More recent industrial applications have adopted advanced architectures such as MMoE (Multi-gate Mixture-of-Experts)~\cite{ma2018modeling} and PLE (Progressive Layered Extraction)~\cite{ple2020recsys} to mitigate negative transfer and capture task-specific nuances.

\noindent\textbf{Multi-Task Fusion and Reinforcement Learning.} 
MTF serves as the critical bridge between multi-objective predictions and the final ranking decision. 
Traditional MTF methods typically rely on heuristic or black-box optimization techniques, such as Bayesian Optimization~\cite{bayesian} and the Cross-Entropy Method~\cite{cem}, to search for optimal fusion weights. 
Although straightforward, these approaches often fail to adapt to dynamic user contexts and predominantly focus on immediate feedback, lacking mechanisms to optimize for long-term satisfaction. To address these limitations, recent research has pivoted toward Reinforcement Learning (RL) based frameworks, wherein a policy dynamically generates personalized fusion weights to maximize cumulative user engagement over time.
BatchRL-MTF~\cite{batchrl_mtf} pioneered this direction by integrating offline batch RL with conservative online exploration. 
RLUR~\cite{rlur} and AURO~\cite{auro} advanced the paradigm by explicitly modeling user retention and handling environment non-stationarity through state abstraction. 
More recently, generative paradigms~\cite{zhao2023user,liu2024sequential,liu2024modeling} have reformulated retention optimization as supervised learning conditioned on return-to-go or probabilistic flows. 
Concurrently, other works have focused on refining credit assignment via future impact decomposition~\cite{wang2024future}, addressing reward sparsity through satisfaction-based shaping~\cite{christakopoulou2022reward}, leveraging human preferences to reinforce long-term engagement~\cite{xue2023prefrec}, enabling fine-grained user-oriented exploration through UOEP~\cite{zhang2024reinforcing}, and uncovering behavioral drivers via interpretable retention modeling with rationale learning~\cite{ding2023interpretable}.

\section{Preliminaries}
\subsection{Problem Formulation}
\label{sec:problem_form}

We formalize the multi-task fusion problem as a constrained optimization problem over fusion weights. 
Let \( q  \) denote the user-query context which contains user-specific and query-specific features. 
For each context \( q \), the fusion module selects a weight vector 
\( \mathbf{w} = (w_1, \dots, w_k)^\top \in \mathbb{R}^k \) through a context-conditioned policy \( \pi_\theta(q) \). 
For notational simplicity, we use \( \mathbf{w} \) throughout the following formulation, while emphasizing that it is dynamically generated based on the current user-query context. Given a set of \( N \) candidate items, each associated with a non-negative multi-objective score vector 
\( \mathbf{s}_n = (s_{n1}, s_{n2}, \dots, s_{nk}) \in \mathbb{R}^k_{\geq 0} \), 
the scoring function is defined as~\cite{batchrl_mtf}:  
\begin{equation}
    f(\mathbf{s}_n, \mathbf{w}) 
    = \sum_{j=1}^k w_j \log(1 + s_{nj})
    = \mathbf{w}^\top \log(\mathbf{1} + \mathbf{s}_n).
    \label{eq:score_func}
\end{equation}

The logarithmic transformation mitigates scale disparities across objectives. 
For a given context \( q \), the selected weight vector \( \mathbf{w} \) is applied to all candidate items in the corresponding ranking request, ensuring a consistent ranking criterion within the list. 
Meanwhile, since \( \mathbf{w} \) is generated by the context-conditioned policy \( \pi_\theta(q) \), it can dynamically adapt across different users and queries.
Items are then ranked in descending order of \( f(\mathbf{s}_n, \mathbf{w}) \) to produce an item list \( L \). The goal is to learn the optimal $\mathbf{w}$ that maximizes the expected return $\mathbb{E}_{L \sim \pi_\theta(q)} [R(q, L)]$, where \( R(q,L) \) quantifies the reward of list \( L \) under context $q$. This optimization is also subject to practical business and fairness
constraints:
\begin{equation}
    \sum_{j=1}^k w_j = 1, \quad
    w_j^{\min} \leq w_j \leq w_j^{\max}, 
    \quad \forall j \in \{1, \dots, k\},
\end{equation}
where $\sum_{j=1}^k w_j = 1$ enforces normalized weight and  the bounds \( [w_j^{\min}, w_j^{\max}] \) prevent any single objective from dominating or being suppressed.

Directly optimizing over continuous \( \mathbf{w} \) under hard constraints is challenging. Instead, we discretize the feasible region. For each objective \( j \), we partition the interval \( [w_j^{\min}, w_j^{\max}] \) into \( B_j \) bins, yielding a discrete action set $\mathcal{A}_j = \{ a_j^{(1)}, a_j^{(2)}, \dots, a_j^{(B_j)} \}$. The full combinatorial action space is the Cartesian product \( \mathcal{A} = \mathcal{A}_1 \times \cdots \times \mathcal{A}_k \). We further define a feasible discrete action set with a small tolerance \( \xi > 0 \):
\begin{equation}
    \mathcal{A}_{\mathrm{feas}} =
    \left\{
    \mathbf{a} \in \mathcal{A}
    \,\middle|\,
    \left| \sum_{j=1}^k a_j - 1 \right| \leq \xi
    \right\}.
\end{equation}
Finally, our fusion policy \( \pi_\theta(q) \) is implemented as a parameterized categorical distribution over 
\( \mathcal{A}_{\mathrm{feas}} \), conditioned on user-specific and query-specific features. 
%For each context \( q \), the selected discrete action corresponds to the fusion weight vector \( \mathbf{w} \) used for ranking.

%As shown in Fig. 2, the module is conditioned on user-specific and query-specific features, so the fusion weights are context-aware and dynamically adapted. We will formalize the notation in the revised version to make it clearer.

\if
We formalize the multi-task fusion problem as a constrained optimization problem over fusion weights. Given a set of \( N \) candidate items, each associated with a non-negative multi-objective score vector \( \mathbf{s}_n = (s_{n1}, s_{n2}, \dots, s_{nk}) \in \mathbb{R}^k_{\geq 0} \), we seek a weight vector \( \mathbf{w} = (w_1, \dots, w_k)^\top \in \mathbb{R}^k \), shared identically across all items, that defines the scoring function~\cite{batchrl_mtf}:  

\begin{equation}
    f(\mathbf{s}_n, \mathbf{w}) = \sum_{j=1}^k w_j \log(1 + s_{nj}) = \mathbf{w}^\top \log(\mathbf{1} + \mathbf{s}_n).
    \label{eq:score_func}
\end{equation}

The logarithmic transformation mitigates scale disparities across objectives. Items are ranked in descending order of \( f(\mathbf{s}_n, \mathbf{w}) \) to produce an item list \( L \). The goal is to learn the optimal \( \mathbf{w} \) that maximizes the expected return $\mathbb{E}_{L \sim \pi} \big[ R(L) \big]$, where \( \pi \) denotes the ranking policy induced by \( \mathbf{w} \), and \( R(L) \) quantifies rewards for list \( L \). This optimization is also subject to practical business and fairness constraints:  
\begin{equation}
    \sum_{j=1}^k w_j = 1, \quad
w_j^{\min} \leq w_j \leq w_j^{\max}, \quad \forall j \in \{1, \dots, k\},
\end{equation}
where $\sum_{j=1}^k w_j = 1$ enforces normalized weight and  the bounds \( [w_j^{\min}, w_j^{\max}] \) prevent any single objective from dominating or being suppressed. Directly optimizing over continuous \( \mathbf{w} \) under hard constraints is challenging. Instead, we discretize the feasible region. For each objective \( j \), we partition the interval \( [w_j^{\min}, w_j^{\max}] \) into \( B_j \) bins, yielding a discrete action set $\mathcal{A}_j = \{ a_j^{(1)}, a_j^{(2)}, \dots, a_j^{(B_j)} \}$. The full combinatorial action space is the Cartesian product \( \mathcal{A} = \mathcal{A}_1 \times \cdots \times \mathcal{A}_k \). We further define a feasible discrete action set with a small tolerance \( \xi > 0 \):  
\begin{equation}
    \mathcal{A}_{\text{feas}} = \left\{ \mathbf{a} \in \mathcal{A} \,\middle|\, \left| \sum_{j=1}^k a_j -1\right| \leq\xi \right\},
\end{equation}
and construct our policy $\pi_\theta$ as a parameterized categorical distribution over the feasible action space. 
\fi

\subsection{Group-Relative Policy Optimization}

Standard policy optimization in reinforcement learning often adopts a trust-region formulation that maximizes expected reward while constraining policy updates to remain close to a reference policy~\cite{trpo}. This constrained problem is commonly expressed as:
\begin{equation}
\max_{\theta} \; \mathbb{E}_{q \sim \mathcal{D}, a \sim \pi_\theta} \left[ A(q, a) \right] \quad \text{s.t.} \quad \text{KL}\left( \pi_\theta(\cdot|q) \,\|\, \pi_{\theta_\text{old}}(\cdot|q) \right) \leq \delta,
\end{equation}
where $A(q, a)$ denotes the advantage of state-action pairs $(q,a)$ and $\delta$ controls the step size. Proximal Policy Optimization (PPO) ~\cite{ppo} approximates this constrained problem using a clipped surrogate objective that restricts the probability ratio between the current and old policies with critic network to estimate advantages. Group-Relative Policy Optimization (GRPO)~\cite{grpo} adapts the PPO framework but eliminates the critic network entirely, instead estimating advantages through group-wise reward normalization. For each state (e.g. a query context) $q$ sampled from the replay buffer $\mathcal{D}$, the policy generates $G$ candidate actions $\{a_g\}_{g=1}^G$ (e.g., each action corresponds to a fusion weight vector for MTF). The GRPO objective is formulated as:
\begin{align}
\mathcal{J}_{\text{GRPO}}(\theta) &= 
\mathbb{E}_{q \sim \mathcal{D},\{a_g\}_{g=1}^G\sim\pi_{\theta_\text{old}}(q)} \Bigg[
\frac{1}{G} \sum_{g=1}^{G}
\min\Bigg(
\rho_g(\theta) \, A(q, a_g), \\ \nonumber
&\qquad\qquad\qquad\qquad \text{clip}\bigg(\rho_g(\theta), 1 - \epsilon, 1 + \epsilon\bigg) \, A(q, a_g)
\Bigg)
\Bigg],
\end{align}
where $\rho_g(\theta)=\frac{\pi_\theta(a_g|q)}{\pi_{\theta_{\text{old}}}(a_g|q)}$, $\epsilon$ is the clipping threshold.

The \textit{group-relative} advantage is estimated by normalizing each action's reward relative to other candidates within the same group: 
\begin{equation}
    A(q, a_g) = \frac{r_g - \mu^{q}}{\sigma^q},
\end{equation}
where $r_g$ is the reward of action $a_g$, and $\mu^{q}$, $\sigma^{q}$ denote the mean and standard deviation of rewards across the $G$ candidate actions for state $q$. This intra-group normalization measures how much better (or worse) an action performs relative to its alternatives under identical states, significantly reducing variance in policy gradient estimates and accelerating convergence particularly advantageous in large language model training, and has since inspired a diverse family of variants~\cite{gdpo,gspo,gfpo,drgrpo,dapo}.

\section{Method}

\begin{figure*}[htbp]
  \centering
  \includegraphics[width=1.05\linewidth]{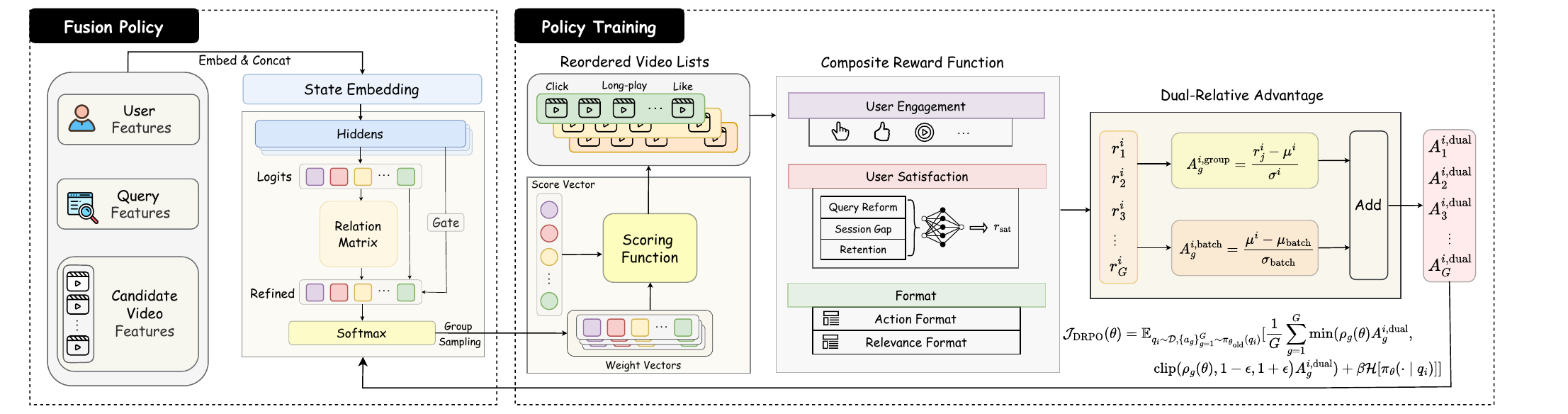} 
  \caption{Overview of the SaFRO framework. After embedding input features into a state, the fusion policy employs relation matrix to output weight distributions. During training, sampled weights (depicted as squares) are fused with their corresponding predicted scores (depicted as circles, e.g. purple for CTR) to generate ranked lists, which are evaluated by a composite reward function, followed by policy updates using dual-relative advantage.}
  \vspace{5pt}
  \label{fig:safro}
\end{figure*}

\subsection{Framework Overview}
In this section, we introduce the overall framework of SaFRO. The framework consists of three core components working in concert: First, a satisfaction-aware reward model constructs query-level satisfaction signals from behavioral proxies that reflects holistic user satisfaction beyond immediate feedbacks. Second, a Dual-Relative Policy Optimization module that enriches policy learning with global information and improves ranking performance. Third, a task relation modeling module captures interdependencies among different objectives through attention-based interactions, enabling context-aware calibration of weight logits. Figure~\ref{fig:safro} shows the overall framework of SaFRO.

\subsection{Satisfaction-Aware Reward Modeling}
\label{sec:sat_model}

\begin{figure}[htbp]
    \centering
    \includegraphics[width=1.0\linewidth]{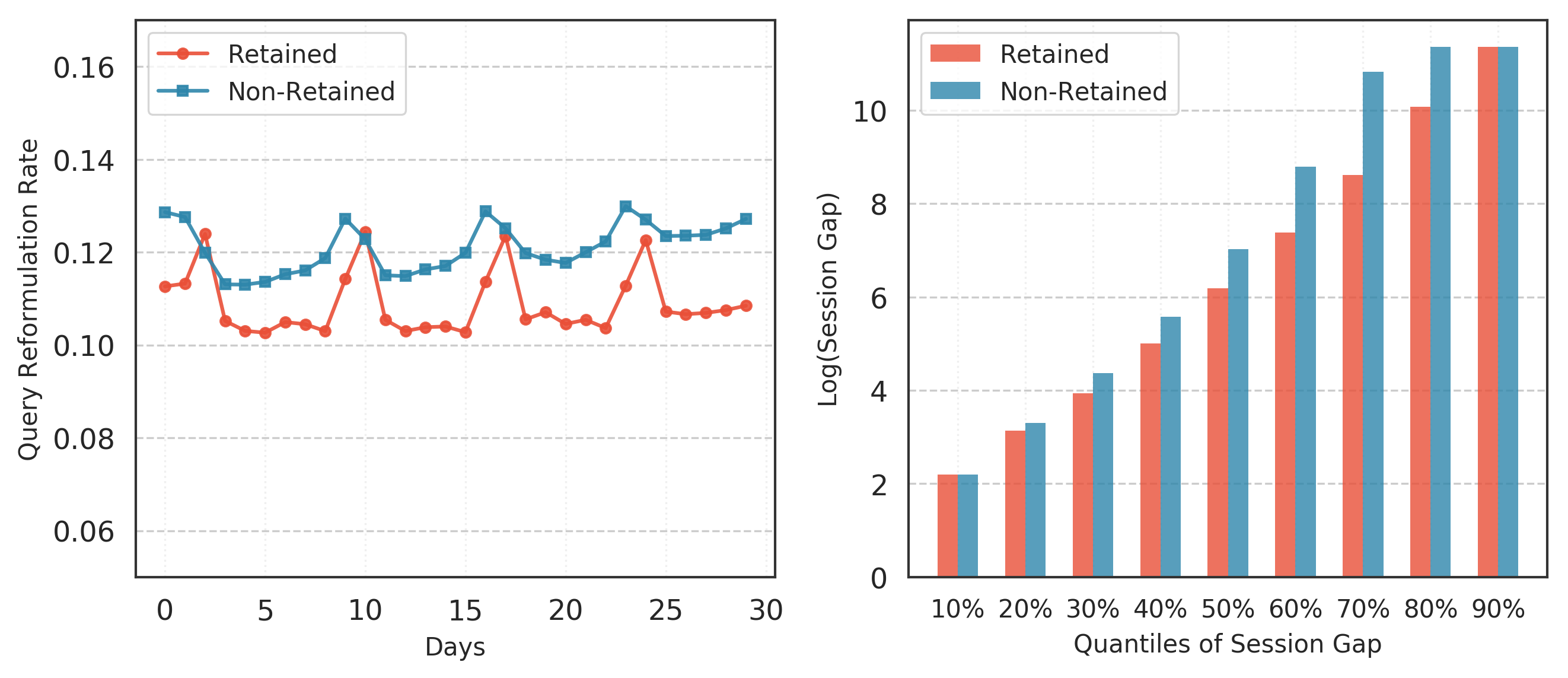}
    \caption{ Comparison of user behavioral patterns by retention status. Left: Daily query reformulation rates observed over a one-month period. Right: Distribution of session gaps (log-scale) across different quantiles.}
    \label{fig:statics_ret}
\end{figure}

Item-level signals such as clicks are abundant but reflect only superficial engagement, whereas long-term retention metrics, although more indicative of genuine satisfaction, are coarse-grained, noisy, and temporally delayed. To bridge this gap, we propose a satisfaction-aware reward model that leverages two query-level behavioral proxies embedded in search logs: 
\begin{itemize}[left=0pt]
    \item \textbf{Query reformulation}, a binary indicator defined as whether a user issues a semantically related follow-up query after viewing the results of the current one.
    \item \textbf{Session gap}, defined as the time interval between the current query and the user’s next query.
\end{itemize}

These two signals jointly capture complementary aspects of user satisfaction at the query level. Query reformulation serves as a strong negative indicator: when users reformulate or modify their queries immediately after viewing results, it typically implies dissatisfaction with the initial retrieval. Conversely, the session gap acts as a continuous, query-level proxy for user retention. Shorter gaps generally correlate with higher engagement levels and a greater likelihood of continued interaction, thereby reflecting a more positive search experience. As illustrated in Figure~\ref{fig:statics_ret}, retained users exhibit a lower query reformulation rate and shorter session gaps compared to non-retained users.

While session gap serves as a valuable proxy for retention, its raw value is confounded by user activity since highly active users naturally exhibit shorter gaps than occasional users. Session gap also exhibits a heavy-tailed distribution with a wide numerical range as shown in Figure~\ref{fig:statics_ret}. To mitigate these biases, we first normalize each user's session gap by their personal historical $\beta$-quantile baseline $\mu_u$ ($\beta=60\%$ in our scenarios), incorporating a small term $\delta$ for stability to obtain \( \tilde{g} = \text{session\_gap} / (\mu_u + \delta) \). We further apply exponential decay transformation to compress the numerical range:
\begin{equation}
    s_{\text{gap}} = \exp\left(-\frac{\tilde{g}}{T}\right) = \exp\left(-\frac{\text{session\_gap}}{(\mu_u + \delta)T}\right),
\end{equation}
where \(T > 0\) is a temperature hyperparameter controlling the decay rate.

Finally, we define a bounded satisfaction reward $r_{\text{sat}} \in [0,1]$ that jointly incorporates query reformulation and retention signals:
\begin{equation}
r_{\text{sat}} = (1 - \mathbb{I}_{\text{reform}}) \cdot \left[ \alpha \cdot \exp\left(-\frac{\tilde{g}}{T}\right) + (1-\alpha) \cdot \mathbb{I}_{\text{ret}} \right],
\end{equation}
where $\mathbb{I}_{\text{reform}}$ indicates query reformulation, $\mathbb{I}_{\text{ret}}$ indicates next-day search retention, defined as whether a user who searches on a given day searches again on the following day. Throughout this paper, retention refers to search retention unless otherwise specified. And $\alpha \in (0,1)$ balances the continuous gap-based signal and the binary retention signal. The multiplicative term $(1 - \mathbb{I}_{\text{reform}})$ ensures that reformulated queries which indicates failed intent fulfillment to receive zero satisfaction reward.

To estimate the satisfaction reward from search interaction logs, we train a list-level reward model \( R_\phi \) that maps the user-query context and ranked list features to a scalar satisfaction score. Formally, given a query \( q_i \) and its associated ranked list \( L_i \), the model takes \( (q_i, L_i) \) as input and predicts \( \hat{r}_i = R_\phi(q_i,L_i) \). Since \( r_{\text{sat}} \) is derived from implicit user behavior and thus inherently noisy, we adopt importance weighting to down-weight low-confidence training samples. The confidence \( c_i \) for each sample is computed based on subsequent user engagement signals within a future time window (e.g., click count and long-play count at time \( T+1 \)), assigning higher confidence to samples with stronger behavioral evidence of satisfaction.

In general, the reward model parameters \( \phi \) are optimized by minimizing the weighted mean squared error:
\begin{equation}
    \min_{\phi} \; \frac{1}{\sum_{i=1}^{N} c_i} \sum_{i=1}^{N} c_i \left( R_\phi(q_i,L_i) - r_{\text{sat}}^{(i)} \right)^2,
\end{equation}
where \( r_{\text{sat}}^{(i)} \) denotes the observed satisfaction reward for the \( i \)-th training sample.

\subsection{Dual-Relative Policy Optimization}
To solve the optimization problem defined in Sec~\ref{sec:problem_form} in a scalable and robust manner, we propose Dual-Relative Policy Optimization (DRPO), which employs dual-relative advantage for policy optimization in discrete action space. 

The dual-relative advantage integrates both \textit{group-relative} and \textit{batch-relative} components. Specifically, we sample a batch of states $\{q_i\}_{i=1}^B$ from the replay buffer $\mathcal{D}$, where $B$ denotes the batch size. For each state $q_i$, we generate a group of $G$ candidate actions $\{ \mathbf{a}_g^{i} \}_{g=1}^G$ and evaluate their corresponding rewards $\{ r_g^{i} \}_{g=1}^G$. The group-relative and batch-relative advantages are computed as:
\begin{equation}
    A_g^{i,\text{group}} = \frac{r_g^i - \mu^i}{\sigma^i}, \quad
     A_g^{i,\text{batch}} = \frac{ \mu^i - \mu_{\text{batch}} }{ \sigma_{\text{batch}} },
\end{equation}
where $\mu^i = \frac{1}{G} \sum_{g=1}^G r_g^i$ and $\sigma^i = \sqrt{ \frac{1}{G} \sum_{g=1}^G (r_g^i - \mu^i)^2 }$ are the mean and standard deviation of rewards within the same group, and $\mu_{\text{batch}} = \frac{1}{B} \sum_{i=1}^B \mu^i$ and $\sigma_{\text{batch}} = \sqrt{ \frac{1}{B} \sum_{i=1}^B (\mu^i - \mu_{\text{batch}})^2}$ represent the mean and standard deviation of group-wise means across the current batch. The dual-relative advantage for action $g$ in query $q_i$ is then obtained by aggregating the two components:
\begin{align}
A_g^{i,\text{dual}} &=A_g^{i,\text{group}} +  A_g^{i,\text{batch}} 
=\frac{r_g^{i} - \mu^{i}}{\sigma^{i}} + \frac{\mu^{i} - \mu_{\text{batch}}}{\sigma_{\text{batch}}}.
\end{align}

\begin{figure}[htbp]
  \centering
  \includegraphics[width=1.0\linewidth]{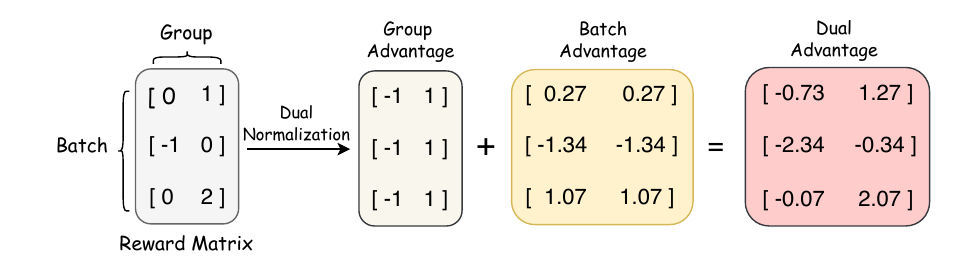} 
  %\vspace{-15pt} 
  \caption{An illustration of dual-relative advantage. The reward matrix is normalized within each group and across the batch based on group-wise means.
}
  %\vspace{-5pt} 
  \label{fig:dual-adv}
\end{figure}

Figure~\ref{fig:dual-adv} presents an example of dual advantage. This dual decomposition can be interpreted as a stratified normalization: the first term measures intra-group relative advantage (how much better action \( g \) is than its peers), while the second term measures inter-batch positional advantage (how well the entire group \( i \) ranks among all groups in the batch). This two-tier structure addresses a key limitation of group-only advantage normalization: while the group-relative term enables fine-grained action ranking within a query, it lacks global context. An action may appear advantageous simply by outperforming poor peers in a low-quality group, leading to misleading updates when the entire group is suboptimal (e.g., due to ambiguous queries or sparse feedback). 

\noindent\textbf{Remark 1.} \textit{Invariant Local Ranking. For a given query $q_i$, the difference in advantage estimation between any two candidate actions $\mathbf{a}_j$ and $\mathbf{a}_k$ depends solely on the group-relative term, i.e., $\Delta A_{j,k}^{i,\text{dual}} = A_j^{i,\text{dual}} - A_k^{i,\text{dual}} = A_j^{i,\text{group}} - A_k^{i,\text{group}}$.}

\noindent\textbf{Remark 2.} \textit{Gradient Decomposition. The policy gradient using dual-relative advantage can be decomposed into two disentangled components: an intra-group ranking term and a query-level importance reweighting term:}
\begin{equation}
\nabla J(\theta) \propto \underbrace{\sum_{i=1}^N \sum_{g=1}^G A_g^{i,\text{group}} \nabla_\theta \log \pi_\theta}_{\text{Term I: Fine-grained Ranking}} 
+ \underbrace{\sum_{i=1}^N C^i \left( \sum_{g=1}^G \nabla_\theta \log \pi_\theta \right)}_{\text{Term II: Batch Reweighting}},
\end{equation}
\textit{where $C^i = (\mu^i - \mu_{\text{batch}}) / \sigma_{\text{batch}}$ represents batch-relative shift scalar.}

\begin{figure}[t]
  \centering
  \includegraphics[width=1.0\linewidth]{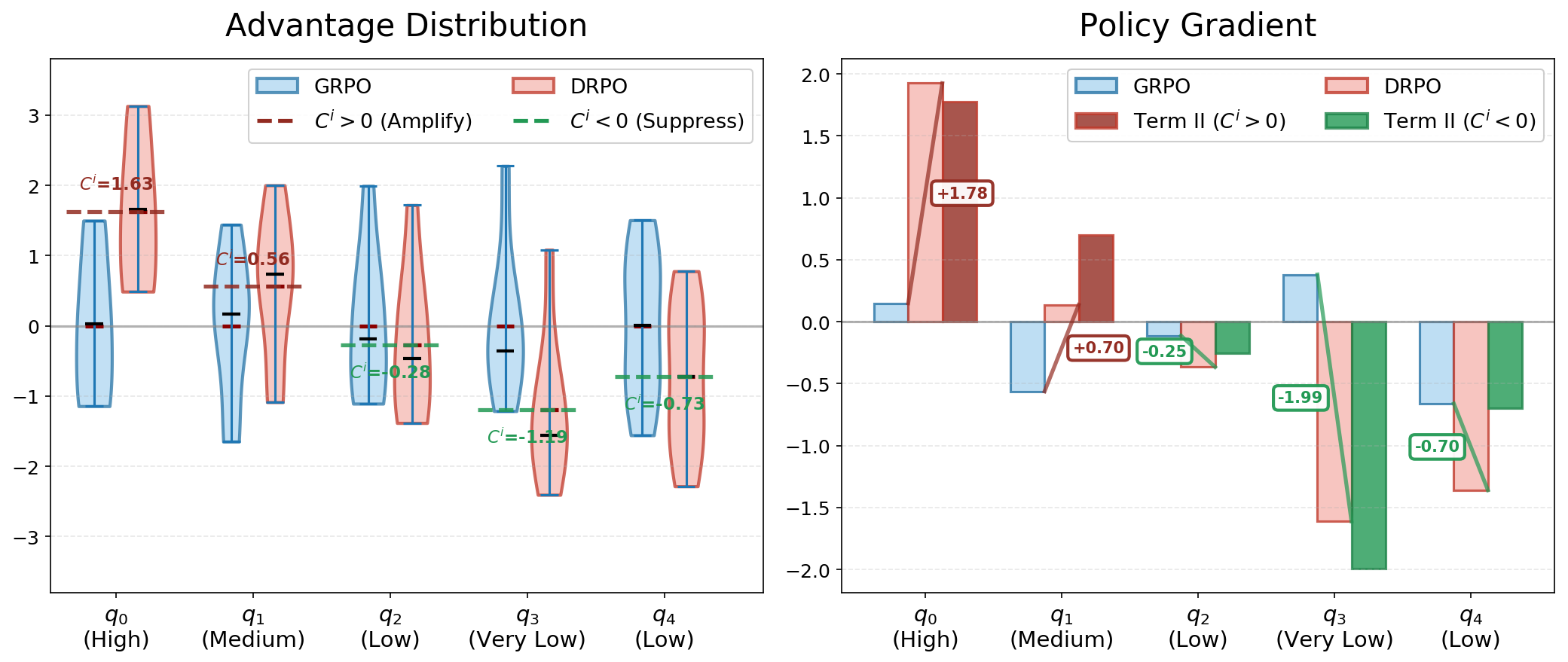} 
  %\vspace{-15pt} 
  \caption{Comparison between GRPO and DRPO. Left: Advantage distribution with red and green dashed lines indicating amplification and suppression, respectively. Right: Policy gradient where arrows and annotations illustrate how Term II transforms GRPO gradients into DRPO gradients.}
  %\vspace{-5pt} 
  \label{fig:dual_final}
\end{figure}

Remark 1 reveals that the gradient direction governing the preference of one action over another is preserved exactly. DRPO does not reduce the model's ability to distinguish between better and worse actions within a specific query. Remark 2 explicates the optimization dynamics of DRPO: Term I preserves the standard GRPO baseline, driving the policy to select the best action relative to the current group mean. Term II functions as a dynamic modulation component, amplifying gradient updates for high-quality queries ($C^i > 0$) while suppressing them for suboptimal ones ($C^i < 0$). We validate this property via simulation on synthetic data with five queries of varying quality levels (from "High" to "Very Low"). As shown in Figure~\ref{fig:dual_final}, DRPO vertically shifts the advantage distribution by $C^i$ relative to GRPO, thereby boosting gradients for high-quality queries ($q_0, q_1$) while attenuating gradients for low-quality queries ($q_2$ to $q_4$). Crucially, the global sum of advantages across the batch remains zero ($\sum_{i,g} A_g^{i,\text{dual}} = 0$), ensuring numerical and optimization stability.

We substitute the dual-relative advantage $A_g^{i,\text{dual}}$ into the GRPO objective and augment the resulting formulation with an entropy bonus term controlled by hyperparameter $\beta$ to encourage exploration:
\begin{align}
\mathcal{J}_{\text{DRPO}}(\theta) &= \mathbb{E}_{q_i \sim \mathcal{D},\{a_g\}_{g=1}^G\sim\pi_{\theta_\text{old}}(q_i)} \Bigg[ \frac{1}{G} \sum_{g=1}^G \min \Bigg( \rho_g(\theta) A_g^{i,\text{dual}}, 
\nonumber
\\ &\text{clip}\big(\rho_g(\theta), 1-\epsilon, 1+\epsilon\big) A_g^{i,\text{dual}} \Bigg) + \beta \mathcal{H}\big[\pi_\theta(\cdot \mid q_i)\big] \Bigg].
\end{align}

To incorporate user immediate feedbacks and long-term satisfaction in DRPO, we design a composite reward function, comprising three components:

\begin{enumerate}[left=0pt]
    \item \textbf{User engagement reward} \( r_{\text{eng}} \):  We adopt a NDCG-style reward to measure user engagement, where the gain of each item is derived from user feedback signals (e.g., click and long-play). This formulation incentivizes the fusion policy to prioritize items with stronger engagement signals, aligning the ranking output with user preference patterns. 
    \item \textbf{User satisfaction reward} \( r_{\text{sat}} \): predicted by the satisfaction model in Sec~\ref{sec:sat_model}, which takes the ranked list as input and outputs a scalar score in \([0,1]\) reflecting holistic user satisfaction.
    \item \textbf{Format reward} \( r_{\text{fmt}} \): enforces operational constraints:     
    \begin{itemize}
        \item \textit{Action-space constraint}: impose a bounded deviation from normalized weight:
        \begin{equation}
        r_{\text{fmt}}^{a} = 
        \begin{cases}
            0, & \left| \sum_{j=1}^k w_j - 1 \right| \leq \xi, \\
            -1, & \text{otherwise}.
        \end{cases}
        \end{equation}

        \item \textit{Relevance constraint}: penalize any ranking in which the top-\(k\) results contain at least one item with relevance below a predefined threshold \( \tau \in (0,1) \):
            \begin{equation}
            r_{\text{fmt}}^{r} = 
            \begin{cases}
            -2, & \exists\, i < k \text{ such that } \text{relevance}(i) < \tau, \\
            0, & \text{otherwise}.
            \end{cases}
            \end{equation}
                \end{itemize}
\end{enumerate}

Finally, We design a composite reward function as:
\begin{equation}
    r = r_{\text{eng}} + r_{\text{sat}} + r_{\text{fmt}}^a + r_{\text{fmt}}^r.
\end{equation}

\subsection{Task Relation Modeling}

Prior works typically assign fusion weights independently for each objective, neglecting the intrinsic interdependencies among tasks~\cite{batchrl_mtf,rlur}. To address this limitation, we propose a Task-Relation-Aware Fusion (TRAF) module that explicitly models pairwise task interactions within the policy network.

Specifically, the policy network takes user features, query features, and candidate item features as input and processes them through Multi-Layer Perceptron (MLP) layers.  After producing the final hidden representation \( \mathbf{h} \) and the base task logits $z_i$ , we first project \( \mathbf{h} \) into a set of task-specific embedding vectors via learnable projection matrix $\mathbf{u}_i = \mathbf{h}\mathbf{W}_u^{(i)}  \in \mathbb{R}^d$. Pairwise interaction scores between tasks are then computed as inner products $e_{ij} = \mathbf{u}_i^\top \mathbf{u}_j$. The refined logit for task $i$ is obtained by adaptively aggregating the base logits using attention weights derived from these interaction scores, combined with a gated residual connection:
\begin{equation}
    \tilde{z}_i = \text{gate}_i \cdot z_i + \sum_{j=1}^{K} \frac{\exp(e_{ij})}{\sum_{k=1}^K \exp(e_{ik})} \cdot z_j.
\end{equation}
These refined logits \( \tilde{\mathbf{z}} = (\tilde{z}_1, \dots, \tilde{z}_K) \) are subsequently passed through a softmax layer to yield the final task weight vector \(\mathbf{w} \sim \text{softmax}(\tilde{\mathbf{z}}) \). By introducing task-relation modeling into logit-level multi-task fusion, TRAF dynamically calibrates each task’s fusion weight according to its learned dependencies with other tasks, enabling adaptive objective balancing rather than relying on fixed weight combinations, while preserving end-to-end differentiability. %This design allows each task’s fusion weight to be dynamically adjusted based on its learned relationships with other tasks while maintaining end-to-end differentiability. 

\section{Experiment}

\begin{table*}[htbp]
\centering
\caption{Overall performance comparison across different methods. We evaluate ranking quality using NDCG metrics across four dimensions (Click, Long-play, Duration, Relevance) and their average. Long-term user satisfaction is evaluated by Satisfaction Score and User Retention. The best results are highlighted in bold.}
\label{tab:main}

\renewcommand{\arraystretch}{1.3}
\begin{tabular}{c|cccc|c|cc}
\Xhline{1pt}
\multirow{2}{*}{\textbf{Method}} &
\multicolumn{5}{c|}{\textbf{NDCG}} &
\multicolumn{2}{c}{\textbf{Satisfaction}} \\
\cline{2-8}
& Click & Long-play & Duration & Relevance & Average &
  Satisfaction Score & User Retention \\
\hline
LambdaRank    & 0.6120 & 0.7513 & 0.6680 & 0.6657 & 0.6743 & 0.7421 & 0.7112 \\
NeuralNDCG    & 0.6306 & 0.7747 & 0.6782 & 0.6904 & 0.6935 & 0.7445 & 0.7183 \\
CEM           & 0.5926 & 0.7168 & 0.6515 & 0.6206 & 0.6454 & 0.7206 & 0.6824 \\
DCN           & 0.6381 & 0.7916 & 0.6829 & 0.7023 & 0.7037 & 0.8422 & 0.8054 \\
DDPG          & 0.6228 & 0.7688 & 0.6671 & 0.6949 & 0.6884 & 0.8132 & 0.7356 \\
TD3           & 0.6280 & 0.7860 & 0.6648 & 0.6947 & 0.6934 & 0.8117 & 0.7321 \\
SAC           & 0.6353 & 0.7864 & 0.6773 & 0.7026 & 0.7004 & 0.8343 & 0.7865 \\
BatchRL-MTF     & 0.6250 & 0.7887 & 0.6645 & 0.6937 & 0.6930 & 0.8202 & 0.7689 \\
RLUR          & 0.6371 & 0.7962 & 0.6838 & 0.6962 & 0.7033 & 0.8398 & 0.8012 \\
AURO          & 0.6384 & 0.7915 & 0.6801 & 0.7059 & 0.7040 & 0.8467 & 0.8026 \\
\hline
\textbf{SaFRO (Ours)} 
& \textbf{0.6431} & \textbf{0.8022} & \textbf{0.6871} & \textbf{0.7085} & \textbf{0.7102}
& \textbf{0.8656} & \textbf{0.8173}  \\
\Xhline{1pt}
\end{tabular}
\end{table*}

In this section, we present a comprehensive empirical evaluation of SaFRO through both extensive offline experiments and large-scale online A/B tests. We begin by outlining the experimental setup, including the industrial dataset and metrics, followed by a comparative analysis against state-of-the-art baselines. Subsequently, we present ablation studies to analyze the contribution of key modules and hyperparameters. Finally, we demonstrate the practical efficacy and business impact of SaFRO through online A/B tests deployed on the Kuaishou search platform.

\subsection{Experimental Setup}
\subsubsection{Dataset} We construct Kuaishou industrial search dataset, which is collected from Kuaishou short-video app with over 400 million daily active users. Each sample corresponds to a user session including user features, query features and candidate item features. The dataset comprises 400 million users, 806 million items and 2.15 billion sessions, which is split into training and testing sets in a 9:1 ratio.

\subsubsection{Metrics} We evaluate different methods in offline experiments on the Kuaishou datasets using NDCG@10 across four item-level signals: click, long-play, duration (user dwell time on each item), and relevance. We also report the satisfaction score predicted by the satisfaction-aware model. Furthermore, we assess user retention rate by building a retention simulator on the same dataset to predict users’ probability of returning the next day. This allows us to assess the long-term effectiveness of each method in an offline setting, where true long-term retention is inherently unobservable and simulator-based evaluation is therefore necessary and widely adopted in industrial practice~\cite{kuaisim,rlur,auro}.
%While a thorough study of simulator fidelity is beyond our current scope, our results show that offline simulated improvements are directionally consistent with real online gains. This strong empirical alignment justifies the use of the simulator as a valid proxy for evaluating long-term user retention.

\subsubsection{Baselines}
We compare SaFRO against a diverse set of baselines, spanning learning-to-rank methods, black-box optimization, value-based RL, and retention-oriented RL algorithms:
\begin{itemize}[left=0pt]
    \item \textbf{LambdaRank}~\cite{lambdarank}: A classic learning-to-rank method that weights training samples based on their impact on the ranking metric.
    \item \textbf{NeuralNDCG}~\cite{neuralndcg}: A differentiable approximation of NDCG, enabling direct optimization of ranking metrics within neural networks.
    \item \textbf{CEM}~\cite{cem}: The Cross-Entropy Method, a gradient-free black-box optimization technique for parameter search.
    \item \textbf{DCN}~\cite{dcn}: A variant that replaces the TRAF module in SaFRO with a Deep \& Cross Network to capture feature interactions.
    \item \textbf{DDPG}~\cite{ddpg}: An Actor-Critic RL algorithm utilizing experience replay for stable learning.
    \item \textbf{TD3}~\cite{td3}: An enhancement over DDPG that incorporates twin Q-networks, delayed policy updates and target smoothing.
    \item \textbf{SAC}~\cite{sac}: A stochastic Actor-Critic algorithm that maximizes both expected reward and policy entropy.
    \item \textbf{BatchRL-MTF}~\cite{batchrl_mtf}: A Batch RL framework designed to optimize user satisfaction.
    \item \textbf{RLUR}~\cite{rlur}: An Actor-Critic RL algorithm specifically tailored for retention optimization.
    \item \textbf{AURO}~\cite{auro}: An Actor-Critic RL algorithm for adaptive user retention optimization.
\end{itemize}

\subsection{Overall Performance}
Table~\ref{tab:main} presents a comprehensive comparison between SaFRO and baseline methods on the Kuaishou dataset. Compared to learning-to-rank approaches like LambdaRank and NeuralNDCG, which strictly prioritize immediate engagement, SaFRO achieves substantial improvements in short-term ranking quality while significantly surpassing them in Satisfaction Score and User Retention. This demonstrates the efficacy of explicitly modeling long-term satisfaction beyond immediate item-level feedback. In contrast, static fusion methods like CEM exhibit the weakest performance due to their inability to adapt to dynamic user intents, highlighting the necessity of SaFRO's adaptive policy-based fusion. 

While DCN achieves competitive performance, it still lags behind SaFRO. The performance gap highlights that replacing the TRAF module with cross-feature interactions is suboptimal; explicitly modeling pairwise task interactions yields superior fusion weights. Furthermore, SaFRO consistently outperforms all RL-based baselines. While value-based methods typically excel in satisfaction metrics but underperform in ranking stability, SaFRO successfully balances both objectives. Notably, compared to retention-centric models (BatchRL-MTF, RLUR, and AURO), SaFRO achieves comparable or superior retention rates while delivering significantly better short-term ranking quality, effectively avoiding the degradation of user experience. Overall, SaFRO achieves the best performance across all evaluation dimensions, validating its effectiveness as a unified framework for satisfaction-aware multi-task fusion.

\begin{figure}[tbp]
  \centering
  \includegraphics[width=1.0\linewidth]{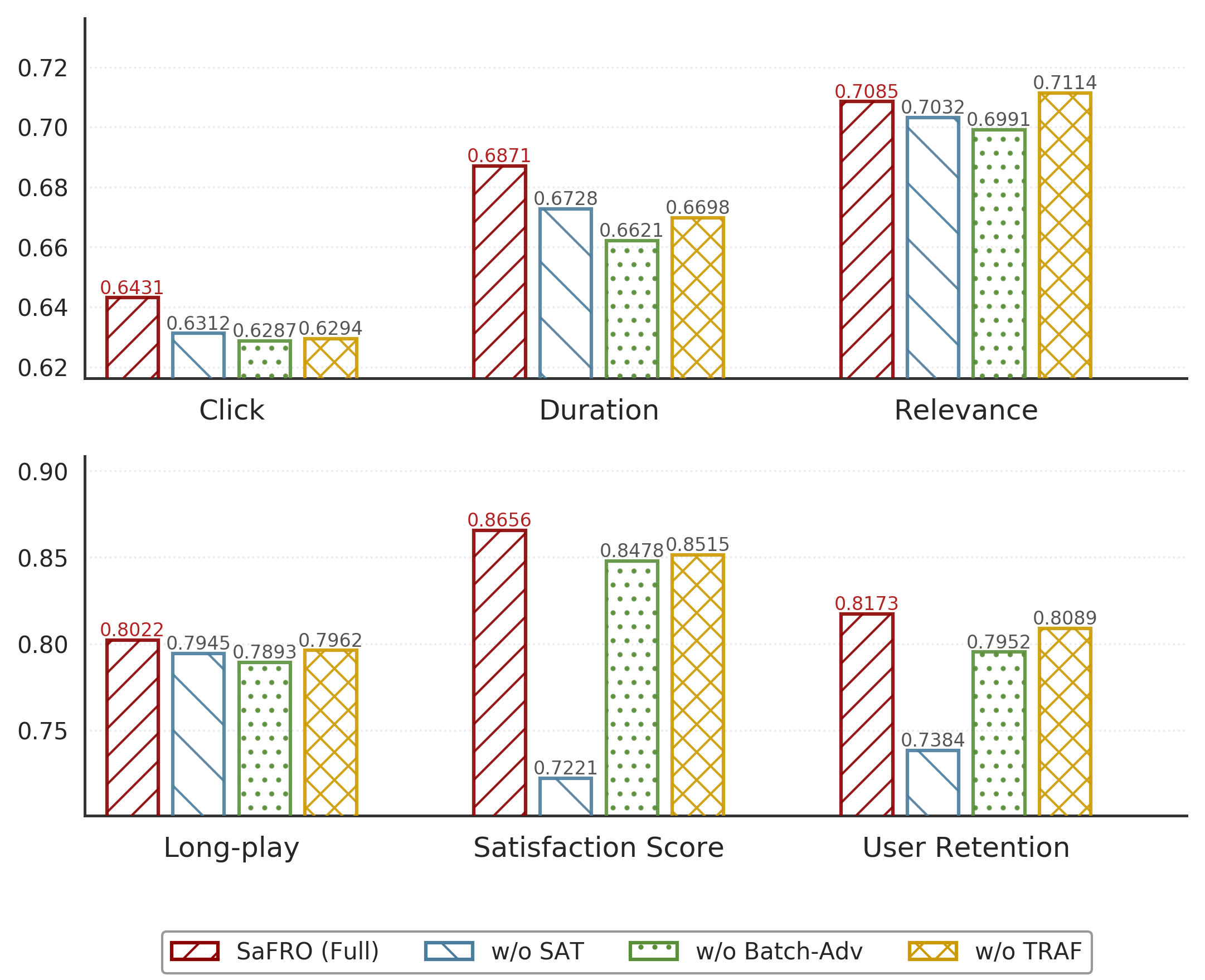} 
  %\vspace{-15pt}
  \caption{Ablation results for key SaFRO modules.}
  \vspace{-5pt}
  \label{fig:ablation}
\end{figure}
\subsection{Ablation Study}
\subsubsection{Module Analysis}
To validate the contribution of each proposed component, we conduct ablation studies by systematically removing or replacing key modules of SaFRO. The results are illustrated in Figure~\ref{fig:ablation}, where we compare the following variants:
\begin{itemize}[left=0pt]
    \item \textbf{w/o SAT}: Removes the satisfaction-aware reward model and relies solely on immediate engagement signals as rewards for policy training.
    \item \textbf{w/o Batch-Adv}: Removes the batch-relative advantage component, using only group-relative advantage estimation (equivalent to standard GRPO).
    \item \textbf{w/o TRAF}: Removes the Task-Relation-Aware Fusion module, outputting task weights without modeling inter-task dependencies.
\end{itemize}
As illustrated in Figure~\ref{fig:ablation}, the ablation study confirms the critical role of each component within the SaFRO framework. First, removing the satisfaction reward model (\textit{w/o SAT}) leads to a sharp deterioration in both Satisfaction Score and User Retention. Notably, this excision also impairs short-term NDCG metrics, suggesting that disregarding delayed satisfaction signals results in myopic policies that fail to capture true user intent. Second, eliminating the batch-relative advantage (\textit{w/o Batch-Adv}) forces the model to rely solely on group normalization, causing significant degradation in both short-term NDCG and long-term metrics, thereby validating the effectiveness of batch advantage normalization. Third, the exclusion of task relation modeling (\textit{w/o TRAF}) consistently degrades performance across all metrics except relevance NDCG, indicating that relevance dominates the other objectives. This demonstrates that modeling inter-task dependencies enables more nuanced trade-offs among competing objectives. Collectively, these results attest to the indispensability of all three core modules and validate the synergistic design of the proposed framework.

\begin{table}[b]
\centering
\caption{Sensitivity analysis of group size $G$ and entropy coefficient $\beta$. The best results are highlighted in bold.}
\label{tab:hyperparam}
\small
\renewcommand{\arraystretch}{1.2}
\begin{tabular}{c|cc|ccc}
\Xhline{1pt}
\multirow{2}{*}{\textbf{Hyperparameter}} & \multicolumn{2}{c|}{\multirow{2}{*}{\textbf{Value}}} & \textbf{NDCG} & \textbf{Satisfaction} & \textbf{User}\\
& & & \textbf{(Average)} & \textbf{Score} & \textbf{Retention}\\
\hline
\multirow{5}{*}{\makecell{\textbf{Group Size} \\ $G$}} 
& \multicolumn{2}{c|}{2}  & 0.6745 & 0.7912 & 0.7380 \\
& \multicolumn{2}{c|}{4}  & 0.6890 & 0.8245 & 0.7795 \\
& \multicolumn{2}{c|}{8}  & 0.6985 & 0.8490 & 0.7960 \\
& \multicolumn{2}{c|}{16} & \textbf{0.7113} & 0.8641 & 0.8168 \\
& \multicolumn{2}{c|}{32} & 0.7102 & \textbf{0.8656} & \textbf{0.8173} \\

\hline
\hline
\multirow{5}{*}{\makecell{\textbf{Entropy} \\ \textbf{Coefficient} \\ $\beta$}} 
& \multicolumn{2}{c|}{0.01} & 0.7065 & 0.8610 & 0.8145 \\
& \multicolumn{2}{c|}{0.05} & \textbf{0.7102} & \textbf{0.8656} & \textbf{0.8173} \\
& \multicolumn{2}{c|}{0.10} & 0.7040 & 0.8605 & 0.8110 \\
& \multicolumn{2}{c|}{0.20} & 0.6855 & 0.8320 & 0.7850 \\
& \multicolumn{2}{c|}{0.30} & 0.6650 & 0.8105 & 0.7425 \\
\Xhline{1pt}
\end{tabular}
\end{table}

\subsubsection{Hyperparameter Analysis on DRPO}

We investigate the sensitivity of our model on two key hyperparameters: the group size $G$ and the entropy coefficient $\beta$. Table~\ref{tab:hyperparam} summarizes the performance variations across different hyperparameter settings.

For the group size $G$, all evaluation metrics exhibit steady improvement as $G$ increases from 2 to 16. The NDCG score reaches its maximum at $G=16$, followed by a slight decline when $G$ is further increased to 32. In contrast, both satisfaction score and user retention continue to rise with larger group sizes, with $G=32$ delivering the highest values. This trend suggests that larger groups enable more effective candidate comparison.

With respect to the entropy coefficient $\beta$, model performance follows a distinct inverted U-shaped pattern. The optimal balance is achieved at $\beta=0.05$, which yields peak performance across all three metrics. A smaller value of $\beta=0.01$ leads to modest performance degradation, indicating inadequate exploration during training. Conversely, larger values such as $\beta=0.30$ impose excessive entropy regularization, reducing NDCG and user retention severely relative to the optimal setting. This deterioration demonstrates that excessive entropy regularization impairs the model's capacity. These results confirm that the configuration $(G=32, \beta=0.05)$ represents a well-balanced choice, achieving superior user-centric outcomes without sacrificing ranking effectiveness.

\subsubsection{Hyperparameter Analysis on Satisfaction-Aware Model} 
\begin{figure}
    \centering
    \includegraphics[width=1.0\linewidth]{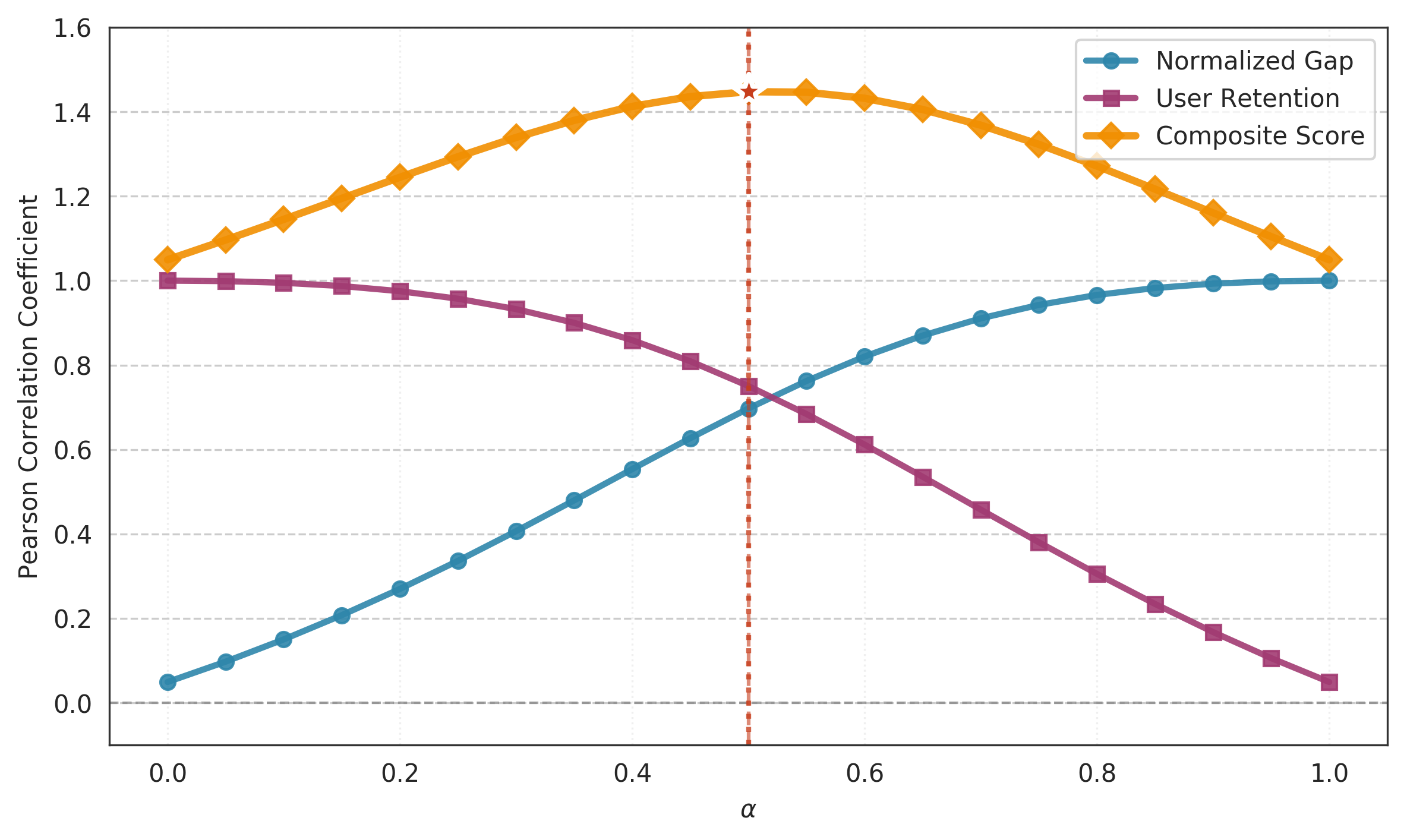}
    %\vspace{-10pt}
    \caption{Pearson correlation analysis between the normalized session gap and user retention. }
    \vspace{-5pt}
    \label{fig:rsat}
\end{figure}

The hyperparameter $\alpha$ governs the trade-off between session gap and user retention. To identify its optimal setting, we conduct a systematic sensitivity analysis by varying $\alpha$ from 0 to 1 with a step size of 0.1. For each configuration, we compute the Pearson correlation coefficients between reward $r_\text{sat}$ and the two constituent terms: the normalized session gap $\exp(-\frac{\tilde{g}}{T})$ and user retention $\mathbb{I}_\text{ret}$. The results are illustrated in Figure~\ref{fig:rsat}. As depicted in the top panel, increasing $\alpha$ strengthens the correlation between $r_\text{sat}$ and the normalized gap term, indicating a greater emphasis on query-level signal. Conversely, the correlation with user retention exhibits a monotonic decrease, reflecting the diminishing influence of the retention component. We define a composite score as the sum of the two correlation coefficients. The bottom panel reveals that this score peaks at $\alpha=0.5$ with a value of approximately 1.45, suggesting that $r_\text{sat}$ achieves optimal alignment with both satisfaction proxies at this point. Consequently, we adopt $\alpha=0.5$ as the default configuration for all experiments.

\subsubsection{Analysis on Implementation Complexity}
%We analyze the implementation complexity of SaFRO from both the training and inference perspectives. During policy training, the dual-relative mechanism introduces only one additional cross-batch normalization step compared with standard GRPO. TRAF further adds a small number of projection and gating matrices, increasing the model size by merely 0.04M parameters, from 0.87M to 0.91M. Accordingly, the training overhead remains marginal: the per-sample training time increases by only 0.53 $\mu$s, from 68.67 $\mu$s to 69.20 $\mu$s, and the computational cost increases by 0.42M FLOPs, from 24.45M to 24.87M. During inference, TRAF introduces negligible additional latency. Specifically, the inference latency increases by only 0.2 ms, from 4.3 ms to 4.5 ms. This indicates that the adaptive fusion mechanism can be applied with minimal impact on online serving efficiency. Overall, SaFRO maintains training and inference costs comparable to those of the baseline, which remains practical for deployment in industrial search systems.

We assess the implementation complexity of SaFRO from both training and inference perspectives. During policy training, the dual-relative mechanism introduces only one additional cross-batch normalization step compared to standard GRPO. TRAF further introduces a limited set of projection and gating matrices, resulting in a modest increase in model size of merely 0.04M parameters (from 0.87M to 0.91M). Consequently, the training overhead remains marginal: per-sample training time increases by only $0.53$ $\mu$s (from $68.67$ $\mu$s to $69.20$ $\mu$s), and computational cost rises by 0.42M FLOPs (from 24.45M to 24.87M). During inference, TRAF introduces negligible additional latency, with inference time increasing by only 0.2 ms (from 4.3 ms to 4.5 ms). This demonstrates that the adaptive fusion mechanism can be integrated with minimal impact on online serving efficiency. Overall, SaFRO maintains training and inference costs comparable to the baseline, making it practical for deployment in industrial search systems.

\subsection{Online Experiment}

\begin{figure}[tbp]
    \centering
    \includegraphics[width=1.0\linewidth]{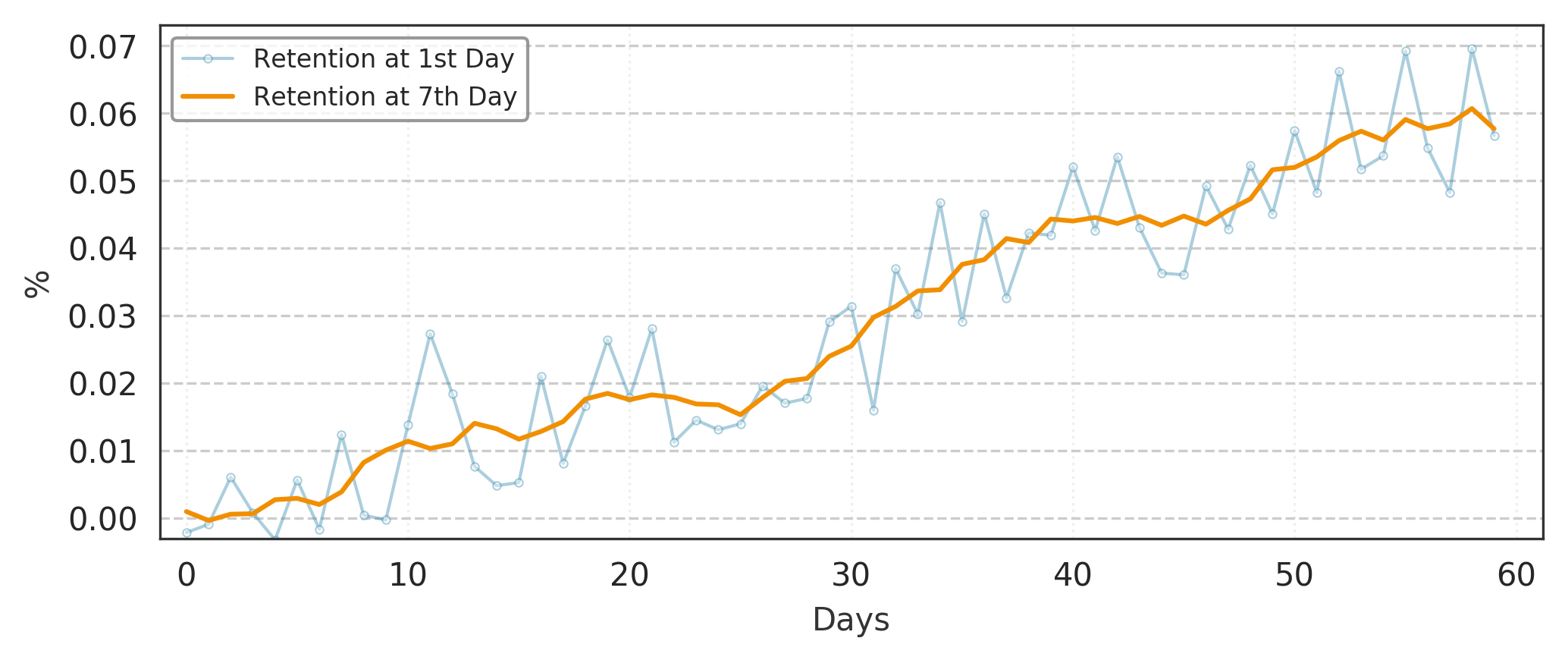}
    %\vspace{-10pt}
    \caption{Retention rate gap of SaFRO over the baseline model in online A/B tests over days (statistically significant with $p<0.05$). }
    %\vspace{-5pt}
    \label{fig:retention_ab}
\end{figure}

To validate the feasibility and effectiveness of SaFRO in a real-world environment, we deployed SaFRO within Kuaishou’s production search system for a large-scale online A/B test. We allocated 10\% of total search traffic to the experimental group over a two-month period, ensuring both statistical significance and broad representativeness across diverse users and query scenarios. The proposed method was evaluated against a competitive DDPG-based baseline optimized for user engagement. As presented in Table~\ref{tab:ab_test}, SaFRO consistently improves immediate user engagement metrics, including CTR, LPR, and watch time, while reducing QRR, suggesting that users are less likely to reformulate queries and thus indicating improved search satisfaction and better intent fulfillment. Furthermore, as illustrated in Figure~\ref{fig:retention_ab}, SaFRO exhibits a sustained upward trend in user retention relative to the baseline over the 60-day period. This continuous divergence empirically verifies SaFRO’s capability to drive long-term user satisfaction and platform loyalty, rather than merely optimizing short-term engagement signals.

%To validate the feasibility and effectiveness of SaFRO in a real-world environment, we deployed SaFRO within Kuaishou’s production search system for a large-scale online A/B test. We allocated 10\% of total search traffic to the experimental group over a two-month period, ensuring both statistical significance and broad representativeness. The proposed method was evaluated against a competitive DDPG-based baseline optimized for user engagement. As presented in Table~\ref{tab:ab_test}, the results confirm that SaFRO significantly improves immediate user engagement metrics. Furthermore, as illustrated in Figure~\ref{fig:retention_ab}, SaFRO exhibits a sustained upward trend in user retention relative to the baseline over the 60-day period. This continuous divergence empirically verifies SaFRO’s capability to drive long-term user satisfaction and platform loyalty. 

\begin{table}[tbp]
\centering
\caption{Online A/B test results ( $p < 0.05$ ). QRR denotes the Query Reformulation Rate, CTR is the Click-Through Rate, and LPR is the Long-Play Ratio.}
\label{tab:ab_test}
\begin{tabular}{c|cccc}
\Xhline{1pt}
\textbf{Metrics} & \textbf{QRR} $\downarrow$ & \textbf{CTR} $\uparrow$ & \textbf{LPR} $\uparrow$ & \textbf{Watch Time} $\uparrow$\\
\hline
\textbf{SaFRO}  & -0.319\% & +0.136\% & +0.495\% & +0.611\%\\
\Xhline{1pt}
\end{tabular}
\end{table}

\section{Conclusion}
In this paper, we present SaFRO, a novel reinforcement learning framework for multi-task fusion in short-video search that explicitly optimizes for long-term user satisfaction. By addressing the inherent challenges of applying RL to search scenarios and the complex attribution of long-term satisfaction, our approach synthesizes a satisfaction-aware reward model, Dual-Relative Policy Optimization, and a Task-Relation-Aware Fusion module into a cohesive system. The satisfaction-aware reward model provides a more comprehensive estimation of user utility by incorporating delayed and holistic satisfaction signals beyond instant clicks or watch time. Dual-Relative Policy Optimization improves the efficiency of policy learning under sparse and delayed feedback, while the Task-Relation-Aware Fusion module adaptively captures dependencies among different objectives, enabling context-sensitive trade-offs across tasks. Extensive offline evaluations and large-scale online A/B tests on the Kuaishou platform demonstrate that SaFRO achieves state-of-the-art performance, establishing a practical paradigm for bridging the gap between immediate ranking signals and sustained user satisfaction. For future work, we aim to extend this framework to other search modalities, leverage large language models for richer state representations, and explore generative architectures to further transcend the limitations of traditional scoring-based methods. 

\bibliographystyle{ACM-Reference-Format}
\bibliography{sample-base}

@String{Computer = "{IEEE} Computer" }

@String{Chelsea = "Chelsea" }

@String{Springer = "Springer-Verlag" }

@article{lambdarank,
  title={Learning to rank with nonsmooth cost functions},
  author={Burges, Christopher and Ragno, Robert and Le, Quoc},
  journal={Advances in neural information processing systems},
  volume={19},
  year={2006}
}

@article{neuralndcg,
  title={Neuralndcg: Direct optimisation of a ranking metric via differentiable relaxation of sorting},
  author={Pobrotyn, Przemys{\l}aw and Bia{\l}obrzeski, Rados{\l}aw},
  journal={arXiv preprint arXiv:2102.07831},
  year={2021}
}

@book{cem,
  title={The cross-entropy method: a unified approach to combinatorial optimization, Monte-Carlo simulation and machine learning},
  author={Rubinstein, Reuven Y and Kroese, Dirk P},
  year={2004},
  publisher={Springer Science \& Business Media}
}

@article{ddpg,
  title={Continuous control with deep reinforcement learning},
  author={Lillicrap, Timothy P and Hunt, Jonathan J and Pritzel, Alexander and Heess, Nicolas and Erez, Tom and Tassa, Yuval and Silver, David and Wierstra, Daan},
  journal={arXiv preprint arXiv:1509.02971},
  year={2015}
}

@inproceedings{td3,
  title={Addressing function approximation error in actor-critic methods},
  author={Fujimoto, Scott and Hoof, Herke and Meger, David},
  booktitle={International conference on machine learning},
  pages={1587--1596},
  year={2018},
  organization={PMLR}
}

@inproceedings{sac,
  title={Soft actor-critic: Off-policy maximum entropy deep reinforcement learning with a stochastic actor},
  author={Haarnoja, Tuomas and Zhou, Aurick and Abbeel, Pieter and Levine, Sergey},
  booktitle={International conference on machine learning},
  pages={1861--1870},
  year={2018},
  organization={Pmlr}
}

@inproceedings{batchrl_mtf,
  title={Multi-task fusion via reinforcement learning for long-term user satisfaction in recommender systems},
  author={Zhang, Qihua and Liu, Junning and Dai, Yuzhuo and Qi, Yiyan and Yuan, Yifan and Zheng, Kunlun and Huang, Fan and Tan, Xianfeng},
  booktitle={Proceedings of the 28th ACM SIGKDD conference on knowledge discovery and data mining},
  pages={4510--4520},
  year={2022}
}

@inproceedings{rlur,
  title={Reinforcing user retention in a billion scale short video recommender system},
  author={Cai, Qingpeng and Liu, Shuchang and Wang, Xueliang and Zuo, Tianyou and Xie, Wentao and Yang, Bin and Zheng, Dong and Jiang, Peng and Gai, Kun},
  booktitle={Companion Proceedings of the ACM Web Conference 2023},
  pages={421--426},
  year={2023}
}

@inproceedings{auro,
  title={AURO: Reinforcement learning for adaptive user retention optimization in recommender systems},
  author={Xue, Zhenghai and Cai, Qingpeng and Yang, Bin and Hu, Lantao and Jiang, Peng and Gai, Kun and An, Bo},
  booktitle={Proceedings of the ACM on Web Conference 2025},
  pages={391--401},
  year={2025}
}

@article{grpo,
  title={Deepseekmath: Pushing the limits of mathematical reasoning in open language models},
  author={Shao, Zhihong and Wang, Peiyi and Zhu, Qihao and Xu, Runxin and Song, Junxiao and Bi, Xiao and Zhang, Haowei and Zhang, Mingchuan and Li, YK and Wu, Yang and others},
  journal={arXiv preprint arXiv:2402.03300},
  year={2024}
}

@inproceedings{trpo,
  title={Trust region policy optimization},
  author={Schulman, John and Levine, Sergey and Abbeel, Pieter and Jordan, Michael and Moritz, Philipp},
  booktitle={International conference on machine learning},
  pages={1889--1897},
  year={2015},
  organization={PMLR}
}

@article{ppo,
  title={Proximal policy optimization algorithms},
  author={Schulman, John and Wolski, Filip and Dhariwal, Prafulla and Radford, Alec and Klimov, Oleg},
  journal={arXiv preprint arXiv:1707.06347},
  year={2017}
}

@inproceedings{bayesian,
  title={On Bayesian methods for seeking the extremum},
  author={Mo{\v{c}}kus, Jonas},
  booktitle={IFIP Technical Conference on Optimization Techniques},
  pages={400--404},
  year={1974},
  organization={Springer}
}

@inproceedings{liu2024sequential,
  title={Sequential Recommendation for Optimizing Both Immediate Feedback and Long-term Retention},
  author={Liu, Ziru and Liu, Shuchang and Zhang, Zijian and Cai, Qingpeng and Zhao, Xiangyu and Zhao, Kesen and Hu, Lantao and Jiang, Peng and Gai, Kun},
  booktitle={Proceedings of the 47th International ACM SIGIR Conference on Research and Development in Information Retrieval},
  pages={1956--1966},
  year={2024},
  organization={ACM}
}

@inproceedings{liu2024modeling,
  title={Modeling User Retention through Generative Flow Networks},
  author={Liu, Ziru and Liu, Shuchang and Yang, Bin and Xue, Zhenghai and Cai, Qingpeng and Zhao, Xiangyu and Zhang, Zijian and Hu, Lantao and Li, Han and Jiang, Peng},
  booktitle={Proceedings of the 30th ACM SIGKDD Conference on Knowledge Discovery and Data Mining},
  pages={5497--5508},
  year={2024},
  organization={ACM}
}

@inproceedings{zhao2023user,
  title={User Retention-oriented Recommendation with Decision Transformer},
  author={Zhao, Kesen and Zou, Lixin and Zhao, Xiangyu and Wang, Maolin and Yin, Dawei},
  booktitle={Proceedings of the ACM Web Conference 2023},
  pages={1141--1149},
  year={2023},
  organization={ACM}
}

@article{christakopoulou2022reward,
  title={Reward Shaping for User Satisfaction in a {REINFORCE} Recommender},
  author={Christakopoulou, Konstantina and Xu, Can and Zhang, Sai and Badam, Sriraj and Potter, Trevor and Li, Daniel and Wan, Hao and Yi, Xinyang and Le, Ya and Berg, Chris and Dixon, Eric Bencomo and Chi, Ed H and Chen, Minmin},
  journal={arXiv preprint arXiv:2209.15166},
  year={2022}
}

@inproceedings{xue2023prefrec,
  title={PrefRec: Recommender Systems with Human Preferences for Reinforcing Long-term User Engagement},
  author={Xue, Wanqi and Cai, Qingpeng and Xue, Zhenghai and Sun, Shuo and Liu, Shuchang and Zheng, Dong and Jiang, Peng and Gai, Kun and An, Bo},
  booktitle={Proceedings of the 29th ACM SIGKDD Conference on Knowledge Discovery and Data Mining},
  pages={2874--2884},
  year={2023},
  organization={ACM}
}

@inproceedings{wang2024future,
  title={Future Impact Decomposition in Request-level Recommendations},
  author={Wang, Xiaobei and Liu, Shuchang and Wang, Xueliang and Cai, Qingpeng and Hu, Lantao and Li, Han and Jiang, Peng and Xie, Guangming},
  booktitle={Proceedings of the 30th ACM SIGKDD Conference on Knowledge Discovery and Data Mining},
  year={2024},
  organization={ACM}
}

@inproceedings{zhang2024reinforcing,
  title={Reinforcing Long-Term Performance in Recommender Systems with User-Oriented Exploration Policy},
  author={Zhang, Changshuo and Chen, Sirui and Zhang, Xiao and Dai, Sunhao and Yu, Weijie and Xu, Jun},
  booktitle={Proceedings of the 47th International ACM SIGIR Conference on Research and Development in Information Retrieval},
  pages={1850--1860},
  year={2024},
  organization={ACM}
}

@inproceedings{ding2023interpretable,
  title={Interpretable User Retention Modeling in Recommendation},
  author={Ding, Rui and Xie, Ruobing and Hao, Xiaobo and Yang, Xiaochun and Ge, Kaikai and Zhang, Xu and Zhou, Jie and Lin, Leyu},
  booktitle={Proceedings of the 17th ACM Conference on Recommender Systems},
  pages={702--712},
  year={2023},
  organization={ACM}
}

@inproceedings{chen2018gradnorm,
  title={Gradnorm: Gradient normalization for adaptive loss balancing in deep multitask networks},
  author={Chen, Zhao and Badrinarayanan, Vijay and Lee, Chen-Yu and Rabinovich, Andrew},
  booktitle={International conference on machine learning},
  pages={794--803},
  year={2018},
  organization={PMLR}
}

@inproceedings{su2024stem,
  title={STEM: unleashing the power of embeddings for multi-task recommendation},
  author={Su, Liangcai and Pan, Junwei and Wang, Ximei and Xiao, Xi and Quan, Shijie and Chen, Xihua and Jiang, Jie},
  booktitle={Proceedings of the AAAI conference on artificial intelligence},
  volume={38},
  number={8},
  pages={9002--9010},
  year={2024}
}

@article{yu2020gradient,
  title={Gradient surgery for multi-task learning},
  author={Yu, Tianhe and Kumar, Saurabh and Gupta, Abhishek and Levine, Sergey and Hausman, Karol and Finn, Chelsea},
  journal={Advances in neural information processing systems},
  volume={33},
  pages={5824--5836},
  year={2020}
}

@article{pei2019value,
  title={Value-Aware Recommendation Based on Reinforced Profit Maximization in E-commerce Systems. arXiv preprint (2019)},
  author={Pei, Changhua and Yang, Xinru and Cui, Qing and Lin, Xiao and Sun, Fei and Jiang, Peng and Ou, Wenwu and Zhang, Yongfeng},
  journal={arXiv preprint arXiv:1902.00851},
  year={2019}
}

@inproceedings{chen2024cache,
  title={Cache-Aware Reinforcement Learning in Large-Scale Recommender Systems},
  author={Chen, Xiaoshuang and Zhang, Gengrui and Wang, Yao and Wu, Yulin and Su, Shuo and Zhan, Kaiqiao and Wang, Ben},
  booktitle={Companion Proceedings of the ACM Web Conference 2024},
  pages={284--291},
  year={2024}
}

@inproceedings{zhang2024unex,
  title={UNEX-RL: reinforcing long-term rewards in multi-stage recommender systems with unidirectional execution},
  author={Zhang, Gengrui and Wang, Yao and Chen, Xiaoshuang and Qian, Hongyi and Zhan, Kaiqiao and Wang, Ben},
  booktitle={Proceedings of the AAAI Conference on Artificial Intelligence},
  volume={38},
  number={8},
  pages={9305--9313},
  year={2024}
}

@book{sutton1998reinforcement,
  title={Reinforcement learning: An introduction},
  author={Sutton, Richard S and Barto, Andrew G and others},
  volume={1},
  number={1},
  year={1998},
  publisher={MIT press Cambridge}
}

@incollection{dcn,
  title={Deep \& cross network for ad click predictions},
  author={Wang, Ruoxi and Fu, Bin and Fu, Gang and Wang, Mingliang},
  booktitle={Proceedings of the ADKDD'17},
  pages={1--7},
  year={2017}
}

@inproceedings{mtl_momma2022,
  author    = {Michinari Momma and Chosen Dong and Yetian Chen},
  title     = {Multi-objective ranking with directions of preferences},
  booktitle = {Proceedings of the ACM SIGIR Workshop on eCommerce},
  year      = {2022},
  pages     = {1--5},
  publisher = {ACM}
}

@inproceedings{cao2020ranking,
  author    = {Xuezhi Cao and Sheng Zhu and Biao Tang and Rui Xie and Fuzheng Zhang and Zhongyuan Wang},
  title     = {Ranking with Deep Multi-Objective Learning},
  booktitle = {Proceedings of the Deep Learning Practice for Industrial Problems},
  year      = {2020},
  pages     = {1--4},
  doi       = {10.1145/3394486.3403288}
}

@article{caruana1997multitask,
  author    = {Rich Caruana},
  title     = {Multitask Learning},
  journal   = {Machine Learning},
  volume    = {28},
  number    = {1},
  pages     = {41--75},
  year      = {1997},
  publisher = {Springer}
}

@inproceedings{misra2016cross,
  author    = {Ishan Misra and Abhinav Shrivastava and Abhinav Gupta and Martial Hebert},
  title     = {Cross-Stitch Networks for Multi-task Learning},
  booktitle = {Proceedings of the IEEE Conference on Computer Vision and Pattern Recognition},
  year      = {2016},
  pages     = {3994--4003}
}

@inproceedings{ma2018modeling,
  author    = {Jiaqi Ma and Zhe Zhao and Xinyang Yi and Jilin Chen and Lichan Hong and Ed H. Chi},
  title     = {Modeling Task Relationships in Multi-task Learning with Multi-gate Mixture-of-Experts},
  booktitle = {Proceedings of the 24th ACM SIGKDD International Conference on Knowledge Discovery \& Data Mining},
  year      = {2018},
  pages     = {1930--1939},
  doi       = {10.1145/3219819.3220007}
}

@inproceedings{ple2020recsys,
  author    = {Hongyan Tang and Junning Liu and Ming Zhao and Xudong Gong},
  title     = {Progressive Layered Extraction (PLE): A Novel Multi-Task Learning (MTL) Model for Personalized Recommendations},
  booktitle = {Proceedings of the 14th ACM Conference on Recommender Systems},
  year      = {2020},
  pages     = {269--278},
  doi       = {10.1145/3383313.3412236}
}

@article{gdpo,
  title={GDPO: Group reward-Decoupled Normalization Policy Optimization for Multi-reward RL Optimization},
  author={Liu, Shih-Yang and Dong, Xin and Lu, Ximing and Diao, Shizhe and Belcak, Peter and Liu, Mingjie and Chen, Min-Hung and Yin, Hongxu and Wang, Yu-Chiang Frank and Cheng, Kwang-Ting and others},
  journal={arXiv preprint arXiv:2601.05242},
  year={2026}
}

@article{drgrpo,
  title={Understanding r1-zero-like training: A critical perspective},
  author={Liu, Zichen and Chen, Changyu and Li, Wenjun and Qi, Penghui and Pang, Tianyu and Du, Chao and Lee, Wee Sun and Lin, Min},
  journal={arXiv preprint arXiv:2503.20783},
  year={2025}
}

@article{gspo,
  title={Group sequence policy optimization},
  author={Zheng, Chujie and Liu, Shixuan and Li, Mingze and Chen, Xiong-Hui and Yu, Bowen and Gao, Chang and Dang, Kai and Liu, Yuqiong and Men, Rui and Yang, An and others},
  journal={arXiv preprint arXiv:2507.18071},
  year={2025}
}

@article{dapo,
  title={Dapo: An open-source llm reinforcement learning system at scale},
  author={Yu, Qiying and Zhang, Zheng and Zhu, Ruofei and Yuan, Yufeng and Zuo, Xiaochen and Yue, Yu and Dai, Weinan and Fan, Tiantian and Liu, Gaohong and Liu, Lingjun and others},
  journal={arXiv preprint arXiv:2503.14476},
  year={2025}
}

@article{gfpo,
  title={Sample more to think less: Group filtered policy optimization for concise reasoning},
  author={Shrivastava, Vaishnavi and Awadallah, Ahmed and Balachandran, Vidhisha and Garg, Shivam and Behl, Harkirat and Papailiopoulos, Dimitris},
  journal={arXiv preprint arXiv:2508.09726},
  year={2025}
}

@article{kuaisim,
  title={KuaiSim: A comprehensive simulator for recommender systems},
  author={Zhao, Kesen and Liu, Shuchang and Cai, Qingpeng and Zhao, Xiangyu and Liu, Ziru and Zheng, Dong and Jiang, Peng and Gai, Kun},
  journal={Advances in Neural Information Processing Systems},
  volume={36},
  pages={44880--44897},
  year={2023}
}

\appendix

\end{document}